\documentclass[12pt]{article}
\usepackage{amsfonts}
\usepackage{graphicx}
\usepackage[utf8]{inputenc}
\usepackage{amsmath,amssymb}
\baselineskip %
\advance %
\textheight by %
\topskip %
\makeatletter \@addtoreset{equation}{section}

\makeatother
\def\Z{\mathbb Z}

\def\R{\mathbb R}
\def\K{{\rm \bf K}}

\newcommand{\half}{{\scriptstyle{\frac{1}{2}}}}

\def\be{\begin{equation}}
\def\ee{\end{equation}}

\def\bea{\begin{eqnarray}}
\def\eea{\end{eqnarray}}

\def\ben{\begin{displaymath}}
\def\een{\end{displaymath}}
\def\ba{\begin{array}{c}}
\def\bal{\begin{array}{l}}
\def\ea{\end{array}}

\def\sn{\mathrm{sn}}
\def\cn{\mathrm{cn}}
\def\dn{\mathrm{dn}}

\def\Z{\mathbb Z}
\def\C{\mathbb C}
\def\R{\mathbb R}

\def\cH{{\mathcal{H}}}

\def\cD{{{\mathcal{D}}}}
\def\cB{{{\mathcal{B}}}}

\def\ii{{\mathbf{i}}}
\def\te{{\tilde{\epsilon}}}
\def\he{{\hat{e}}}
\def\bI{{\mathbb{I}}}
\def\Si{{\Sigma}}
\def\csch{{\rm csch}}
\begin{document}
\title{Extended supersymmetry of the
self-isospectral crystalline and soliton chains}
\author{\textsf{Adri\'an Arancibia and Mikhail S. Plyushchay}
\\
{\small \textit{Departamento de F\'{\i}sica, Universidad de Santiago
de Chile, Casilla 307, Santiago 2,
Chile}}\\
}
\date{}
\maketitle
\begin{abstract}
We study supersymmetric structure of the self-isospectral
crystalline chains formed by $N$ copies of the mutually displaced
one-gap Lam\'e systems. It is generated by the $N(N-1)$ integrals of
motion which are the first order matrix differential operators, by
the same number of the nontrivial second order integrals, and by the
$N$ third order Lax integrals. We show that the structure admits
distinct alternatives for a grading operator, and in dependence on
its choice one of the third order matrix integrals plays either the
role of the bosonic central charge or the role of the fermionic
supercharge to be a square root of the spectral polynomial. Yet
another peculiarity is that the set of all the second order
integrals of motion generates a nonlinear sub-superalgebra. We also
investigate the associated self-isospectral soliton chains, and
discuss possible physical applications of the unusual extended
supersymmetry.
\end{abstract}

\section{Introduction}

Quantum periodic finite-gap systems are associated with completely
integrable models. Any such a system is characterized by a proper
nontrivial integral of motion that is a higher order Lax operator
$\mathcal{P}$ \cite{finitegap}. In the infinite period limit they
transform into reflectionless (soliton) systems.  Their band
structure  is effectively encrypted  in a Burchnall-Chaundy operator
identity \cite{BCh,Kri} that can be presented in a form of a
supersymmetric-like relation \cite{CP1}
\begin{equation}\label{hidnsusy}
    \{\mathcal{P},\mathcal{P}\}=2P(H),
\end{equation}
where $P(H)$ is a spectral polynomial and $H$ is a
Hamiltonian
\footnote{Because of a higher order nature of the Lax integral, the
indicated relation formally corresponds to a higher-derivative
generalization of the supersymmetric quantum mechanics \cite{AIS},
whose construction naturally  arises after truncation of the
parasupersymmetric quantum mechanics \cite{RubSpi}. Nonlinear
supersymmetry of an extended system with higher order superscharges
can be related to the Crum-Darboux transformations
\cite{AIN,BagSam,FerSUSYn,KP1}  in the way like the usual
supersymmetric quantum mechanics is related to the first order
Darboux transformations \cite{MatSal}. In the unextended case,
however, relation  (\ref{hidnsusy}) has a nature of a \emph{hidden}
nonlinear supersymmetry~\cite{MP1}, in which the $\Z_2$-grading is
provided by a reflection operator \cite{boso2} that identifies the
parity-odd Lax integral as a fermionic supercharge \cite{CP1}. In
the extended case of the chains of finite-gap systems we study here,
the Crum-Darboux and hidden supersymmetric structures
 naturally meet. For the earlier treatment of finite-gap systems in
supersymmetric  and related contexts  see also
\cite{BraMac,SaxBish,VesShab,DunFei,FNN,Sams,AnSok}.}.

Because of the presence of the nontrivial integral $\mathcal{P}$, a
supersymmetric extension of finite-gap systems has a more rich
structure in comparison with that of a usual $N=2$ supersymmetric
quantum mechanics.  An example of a physical system with such an
unusual supersymmetry is provided by a model of a non-relativistic
electron in periodic magnetic and electric fields \cite{CJNP}. In a
generic case of the $N=2$ superextension of an $n$-gap system by
means of a Crum-Darboux transformation of the order $m$, $1\leq
m<2n+1$, where $2n+1$ is a number of band edge states,
supersymmetric finite-gap system is characterized by two further
supercharges of the order $(2n+1-m)$ \cite{CJNP}. The
anti-commutator of  supercharges of the orders $m$ and $(2n+1-m)$
generates the Lax integral of the order $2n+1$. Another important
peculiarity is that such superextended systems admit various choices
for the grading operator. For some of them, matrix diagonal Lax
integral takes a role of one of the supercharges that annihilates
all the $2(2n+1)$ band edge states of the extended system.  Unlike
the anti-diagonal supercharges of the orders $m$ and $(2n+1-m$), it
also distinguishes the left- and right- moving Bloch modes inside
the allowed bands \cite{PAN}. This is similar  to a role played by a
momentum operator for a free particle that can be treated as a
zero-gap system.

Sometimes, a supersymmetric partner happens to be just a spatially
displaced initial finite-gap periodic or non-periodic system. The
Witten index takes then a zero value for such a
\emph{self-isospectral} system \cite{DunFei} even in the case of
unbroken supersymmetry \cite{BraMac,SaxBish,CJPJA}.

 Recently, it was observed \cite{PAN} that a phase transition
between the kink-antikink and kink crystalline phases in the
Gross-Neveu model \cite{GN,Thies,Basar,Ebert,CarNick} is accompanied
by the structural changes in the associated supersymmetric
self-isospectral \emph{one-gap} periodic Lam\'e system. Any of the
two first order supercharges of the latter can be taken as the
Bogoliubov-de Gennes Hamiltonian in the Andreev approximation, in
which superpotential is identified with a gap function (a condensate
field). The first order supercharges are constructed from the
Darboux displacement generators of the associated second order
Schr\"odinger system. Yet another peculiarity of such a system is
that the first order Bogoliubov-de Gennes Hamiltonian possesses its
own, exotic hidden nonlinear supersymmetry \cite{PAN}.  A certain
infinite period limit applied to the one-gap self-isospectral system
reproduces \emph{either} the supersymmetric structure of the Dashen,
Hasslacher, and Neveu kink-antikink baryons \cite{DaHN} as a Darboux
dressed form of a free massive Dirac particle \cite{PNPRD},
\emph{or} superextended version of the Callan-Coleman-Gross-Zee kink
solution \cite{DaHN,Gross} of the Gross-Neveu model \cite{PAN}.

A pair of the second order supercharges of the $N=2$ superextended
one-gap Lam\'e system is generated by a sequence of the two first
order Darboux displacements, while the Lax operators of the mutually
shifted Lam\'e subsystems are produced  by closed third order
Crum-Darboux loops. The second order supercharges and the third
order Lax integrals include a dependence on an auxiliary, virtual
displacement parameter. It is natural to try to extend the model by
taking $N>2$ mutually displaced copies of Lam\'e systems. This
produces then the question:
\begin{itemize}
\item What supersymmetric
structure will be generated in such an extended Lam\'e system by
considering the higher order unclosed ($m>2$) and closed ($m>3$)
sequences of Darboux displacements?
\end{itemize}
In the present paper we study the  supersymmetric structure of the
self-isospectral crystalline chain of mutually displaced one-gap
Lam\'e systems, and investigate in the same context the associated
 self-isospectral  non-periodic (soliton) chains of reflectionless
P\"oschl-Teller systems. \vskip0.07cm

The paper is organized as follows. In the next section we construct
the first order Darboux displacement generators for the one-gap
Lam\'e system. In section 3 we apply them to construct
self-isospectral chains of Lam\'e systems, and investigate the
higher order generalizations of the Darboux displacement generators.
In section 4 we study a general structure of the exotic nonlinear
supersymmetry of the self-isospectral crystalline $N$-term chain.
Section 5 is devoted to the discussion of alternative choices for
the $\Z_2$-grading operator admitted by such an extended
supersymmetric structure. In section 6 general theory is illustrated
by an example of the $N=3$ crystalline chain. In section 7 we
consider the infinite period limit that produces a self-isospectral
non-periodic chain and its supersymmetric structure. In the last
section we conclude with a discussion of possible physical
applications of the revealed unusual nonlinear supersymmetry.

\section{Darboux displacement generators: periodic case}

Consider a  one-dimensional self-adjoint Schr\"odinger Hamiltonian
$H(x)=-\frac{d^2}{dx^2}+V(x)$ with a periodic potential $V(x)$.
Require that the system admits a family of the first order Darboux
displacement generators
$\mathcal{D}(x;\lambda)=\frac{d}{dx}+\varphi(x;\lambda)$,
\begin{equation}\label{Daral}
    \mathcal{D}(x;\lambda)H(x)=H(x+\lambda)\mathcal{D}(x;\lambda),
\end{equation}
which depends on a continuous parameter $\lambda$. Then it can be
shown that  $V(x)$ has to be a one-gap Lam\'e potential
\cite{SaxBish,Sams}. The Jacobi form of the one-gap Lam\'e system is
\cite{WW}
\begin{equation}\label{HLame}
    H(x)=-\frac{d^2}{dx^2}+2k^2\sn^2\,x-k^2\,.
\end{equation}
A modular parameter $k$,  $0<k<1$, fixes a real, $2\K$, and
imaginary, $2\ii \K'$, periods of the potential. Here and in what
follows we do not indicate explicitly the dependence of elliptic and
related functions on $k$, and use a notation $\ii=\sqrt{-1}$ for
imaginary unit. The chosen value of the additive constant fixes the
level of the lower edge of the valence band to be zero, and the
one-gap spectrum of (\ref{HLame}) is $\sigma(H)= [0,k'^2]\cup
[1,\infty)$, where $0<k'<1$ is a complementary modular parameter,
$k'^2=1-k^2$. The infinite period limit ($k\rightarrow 1$
$\Rightarrow$ $2\K\rightarrow \infty$, $2\ii\K'\rightarrow \ii\pi$,
$\sn\,x\rightarrow \tanh x$) of (\ref{HLame}) corresponds to a
reflectionless P\"oschl-Teller system with one bound state in the
spectrum, while in another limit $k\rightarrow 0$,  (\ref{HLame})
reduces to a free particle system.

To construct a  one-parametric Darboux displacement generator, we
discuss shortly some properties of Lam\'e system (\ref{HLame}).
Solutions of the stationary equation $H\Psi(x)=E\Psi(x)$ are given
by the Bloch functions of the form
\begin{equation}\label{Psipm}
    \Psi_\pm^{\alpha}(x)=\frac{{\rm H}(x\pm \alpha)}{\Theta(x)}
    \exp\left[
    \mp x {\rm Z}(\alpha)\right]\,,\qquad
    E=\dn^2\alpha,
\end{equation}
where ${\rm H}$, $\Theta$ and ${\rm Z}$ are the Eta, Theta and Zeta
Jacobi functions \cite{WW,Lawden}. Under translation for the period,
they transform as
\begin{equation}\label{Bloch}
    \Psi_\pm^{\alpha}(x+2\K)=
    \exp[\mp \ii 2\K\kappa(\alpha)]\Psi_\pm^{\alpha}(x)\,,\quad
    \kappa(\alpha)=\frac{\pi}{2\K}-\ii{\rm Z}(\alpha)\,,
\end{equation}
where $ \kappa(\alpha)$ is a quasi-momentum. Energy $E$ is given
here as a function $E(\alpha)=\dn^2\alpha$ of a complex parameter
$\alpha$. This is an elliptic function with the same modular
parameter $k$, and its period parallelogram in complex plane
$\alpha\in\C$  is a rectangle with vertices in $0$, $2\K$, $2\K+
2\ii\K'$, and $2\ii\K'$. On the border of the indicated period
parallelogram, function $\dn\,\alpha$ takes real or pure imaginary
values, and so, $E$ is real. The vertical sides
$\alpha=\ii\beta+\K$, $0\leq \beta\leq \K'$, and $\alpha=\ii\beta$,
$0\leq \beta< \K'$, correspond, respectively, to the valence, $0\leq
E\leq k'^2$, and the conduction, $1\leq E<\infty$, bands, where the
quasi-momentum $\kappa(\alpha)$ is real. The horizontal sides
$\alpha=\ii\K'+\beta$ and $\alpha=\beta$ with $0< \beta< \K$
correspond to the prohibited bands $-\infty<E<0$ and $k'^2<E<1$,
where $\kappa(\alpha)$ takes complex values. Inside the allowed
bands, (\ref{Psipm}) are the two Bloch modes propagating to the left
(the upper index) and to the right (the lower index). On the edges
of the bands, they reduce to the \emph{standing waves} described  by
the periodic, $\dn\, x=\dn\,(x+2\K)$ ($E=0$), and antiperiodic,
$\cn\,x=-\cn\,(x+2\K)$ ($E=k'^2$) and $\sn\, x=-\sn(x+2\K)$ ($E=1$),
functions.

Like the ground state $\dn\,x$ ($\alpha=\K+ \ii\K'$, $E=0$),
non-physical Hamiltonian eigenstates  in the lower prohibited band
$-\infty<E<0$ are nodeless functions, which can be used to construct
a one-parametric family of the first order Darboux generators. As
$\dn\,(-u)=\dn\,(u+2\K)=-\dn\,(u+2\ii\K')=\dn\,u$, it is convenient
to introduce a notation $\alpha=-2\tau+\ii\K'$, and assume that
$\tau\in\R$ while keeping in mind that $E\rightarrow-\infty$ for
$\tau \rightarrow n\K$, $n\in\Z$. By shifting the argument,
$x\rightarrow x+\tau$, for the wave function (\ref{Psipm}) with the
upper index we get $\Psi^{-2\tau+\ii\K'}_+(x+\tau)=
c(\tau)F(x;\tau)$, where $c(\tau)$ is a nonzero $x$-independent
multiplier, and
\begin{equation}\label{ThF}
    \frac{\Theta(x-\tau)}{\Theta(x+\tau)}
    \exp[x \mathfrak{z}(\tau)]\equiv
    F(x;\tau)\,.
\end{equation}
Here
\begin{equation}\label{ztau}
    \mathfrak{z}(\tau)={\rm Z}(2\tau+\ii\K')
    +\ii\frac{\pi}{2\K}=
    {\rm Z}(2\tau)+
    \frac{\cn\,2\tau\dn\,2\tau}{\sn\,2\tau\,}
\end{equation}
is, up to a factor $-\ii$, a quasi-momentum of the Bloch state
(\ref{ThF}), $\mathfrak{z}(\tau)=-\ii\kappa(-2\tau+\ii\K')$, that is
an \emph{odd function} of $\tau$. A nodeless function $F(x;\tau)$ is
quasi-periodic in $x$, $F(x+2\K;\tau)=\exp[2\K
    \mathfrak{z}(\tau)]F(x;\tau)$, periodic in $\tau$,
$F(x;\tau+2\K)=F(x;\tau)$, and  for $x\neq 0$ it undergoes infinite
jumps from $0$ to $+\infty$ at $\tau=n\K$, $n\in\Z$.  It also
satisfies the relations $F(x;-\tau)=F(-x;\tau)={1}/{F(x;\tau)}$. In
the case of the ground state  $\mathfrak{z}(-\K/2)=0$, and function
(\ref{ThF}) reduces to a periodic function
$(k')^{-1/2}\dn(x-\frac{\K}{2})$.

Let us consider now a first order differential operator
\begin{equation}\label{Ddef}
    \mathcal{D}(x;\tau)=F(x;\tau)\frac{d}{dx}\frac{1}{F(x;\tau)}
    =\frac{d}{dx}-\Delta(x;\tau),\qquad
    \mathcal{D}^\dagger(x;\tau)=-\mathcal{D}(x;-\tau)\,,
\end{equation}
whose zero mode is $F(x;\tau)$, $\mathcal{D}(x;\tau)F(x;\tau)=0$.
Function $\Delta(x;\tau)={F\,{}'(x;\tau)}/{F(x;\tau)}$,
$F\,{}'(x;\tau)=\frac{\partial}{\partial x}F(x;\tau)$, reads
\begin{eqnarray}
    \Delta(x;\tau)&=&
    \mathfrak{z}(\tau)+{\rm Z}(x-\tau)-{\rm Z}(x+\tau)\nonumber\\
    &=&
    \frac{\cn\,2\tau \dn\,2\tau}{\sn\,2\tau} +k^2\sn\,2\tau
    \sn(x-\tau)\sn(x+\tau)\,.\label{Deldef}
\end{eqnarray}
It obeys the Riccati equations
\begin{equation}\label{Ricc}
    \Delta^2(x;\tau)\pm\Delta'(x;\tau)=
    2k^2\sn^2(x\pm \tau)-k^2+\varepsilon(\tau)\,,
\end{equation}
where
\begin{equation}\label{vareps}
    \varepsilon(\tau)=-E(-2\tau+\ii\K')=\cn^22\tau/\sn^22\tau\,.
\end{equation}
Another important property is that the following three-term linear
combination,
\begin{eqnarray}
    &&\Delta(x;\tau)+\Delta(x+\tau+\lambda;\lambda)+\Delta(x+\lambda;-\tau-\lambda)
    \nonumber\\
    &&=\mathfrak{z}(\tau)+
    \mathfrak{z}(\lambda)+\mathfrak{z}(-\tau-\lambda)\equiv
    g(\tau,\lambda)
    \label{DDDx}\,,
\end{eqnarray}
is $x$-independent.  The function $g(\tau,\lambda)$ possesses the
symmetry properties
$g(\tau,\lambda)=g(\lambda,\tau)=g(\tau,-\lambda-\tau)=-
g(-\tau,-\lambda)$ and can be presented in a form
\begin{equation}\label{gcs}
    g(\tau,\lambda)=\frac{1-\cn\,2\tau\,
    \cn\,2\lambda\,\cn\,
    2(\tau+\lambda)}{
    \sn\,2\tau\,\sn\,2\lambda\,\sn\,2(\tau+\lambda)}\,.
\end{equation}
{}From Eqs. (\ref{ztau}) and (\ref{Ddef}) we find a relation
$\Delta(\xi;\tau)=g
    \big(\frac{1}{2}(\xi-\tau-\ii\K');\tau\big)$,
i.e.  by (\ref{DDDx}) $\Delta(\xi;\tau)$ can be presented by a
three-term sum of quasi-momenta taken at three values of arguments
which sum up to the zero. We also will need the identity
\begin{eqnarray}
    &&\Delta'(x+\tau+\lambda;\lambda)-
    \Delta(x+\lambda;\tau+\lambda)\Delta(x+\tau+\lambda;\lambda) +
    g(\tau,\lambda)\Delta(x;\tau)=\nonumber\\
    &&-\frac{1}{2}
    (\Delta^2(x;\tau)+\Delta'(x;\tau)+\delta(\tau))\,,
    \quad {\rm where}\quad
    \delta(\tau)=1+k^2-3\sn^{-2}2\tau\,,
    \label{Dgdel2}
\end{eqnarray}
which follows from (\ref{Ricc}) and (\ref{DDDx}).

By the Riccati equations (\ref{Ricc}), the operators (\ref{Ddef})
factorize the Lam\'e Hamiltonian (\ref{HLame}),
\begin{equation}\label{DDH}
    \mathcal{D}^\dagger(x;\tau)
    \mathcal{D}(x;\tau)=H(x+\tau)+
    \varepsilon(\tau)\,,\qquad
    \mathcal{D}(x;\tau)
    \mathcal{D}^\dagger(x;\tau)=H(x-\tau)+
    \varepsilon(\tau)\,,
\end{equation}
with (\ref{vareps}) playing a role of a factorization constant.

 Note here that the second eigenstate
 $\Psi^{-2\tau +\ii\K'}_-(x+\tau)$
of the shifted Lam\'e Hamiltonian operator (\ref{HLame}) reduces, up
to an inessential $\tau$-dependent multiplier, to
$F(x+2\tau;-\tau)$, that is coherent with a relation $E(-2\tau
+\ii\K')=E(2\tau +\ii\K')$. In correspondence with this observation,
a change $\tau\rightarrow-\tau$ in the first relation from
(\ref{DDH}) and a subsequent shift  $x\rightarrow x+2\tau$ transform
this first factorization into an equivalent form
$\mathcal{D}(x+2\tau;\tau)
    \mathcal{D}^\dagger(x+2\tau;\tau)=H(x+\tau)+
    \varepsilon(\tau)$, that is just the second relation
    from (\ref{DDH}) with the argument $x$ shifted for $2\tau$.

{}From (\ref{DDH}) it follows that (\ref{Ddef}) are the sought for
Darboux displacement generators,
\begin{equation}\label{DDarD}
    \mathcal{D}(x;\tau)H(x+\tau)=H(x-\tau)\mathcal{D}(x;\tau)\,,\qquad
    \mathcal{D}^\dagger(x;\tau)H(x-\tau)=H(x+\tau)\mathcal{D}^\dagger(x;\tau).
\end{equation}
As $\mathcal{D}^\dagger(x;\tau)= -\mathcal{D}(x;-\tau)$, it is
sufficient to consider only the first intertwining relation from
(\ref{DDarD}) while the second follows from it via a simple change
$\tau\rightarrow-\tau$.

\section{Chains of one-gap Lam\'e systems}

In this section we construct higher order generalizations of Darboux
generators, that will lead us naturally to the chains of
Darboux-displaced one-gap Lam\'e systems.

A mutual spatial Darboux displacement between the two systems in
(\ref{DDarD}) is $2\tau$ while their `average coordinate' is $x$.
For generalization it is convenient to characterize each of the two
related systems by its own shift parameter by introducing the
notations
\begin{equation}\label{tauxab}
    \tau_{ab}=\frac{1}{2}(\tau_b-\tau_a)=-\tau_{ba}\,,\qquad
    x_{ab}=x+\frac{1}{2}(\tau_a+\tau_b)=x_{ba}.
\end{equation}
Then  $x_{ab}+\tau_{ab}=x+\tau_b$, $x_{ab}-\tau_{ab}=x+\tau_a$,
and relations (\ref{DDH}), (\ref{DDarD}) can be presented in the
form
\begin{equation}\label{DDHab}
    \mathcal{D}_{ab}\mathcal{D}^\dagger_{ab}=
    -\mathcal{D}_{ab}\mathcal{D}_{ba}=H_a+\varepsilon_{ab},
\end{equation}
\begin{equation}\label{DabH}
    \mathcal{D}_{ab}H_b=H_a\mathcal{D}_{ab},
\end{equation}
where we have introduced further notations
\begin{equation}\label{Dab}
    \mathcal{D}_{ab}=\mathcal{D}(x_{ab};\tau_{ab})=-\mathcal{D}^\dagger_{ba},\qquad
    H_a=H(x+\tau_a),\qquad
    \varepsilon_{ab}=\varepsilon(\tau_{ab})=\varepsilon_{ba}\,,
\end{equation}
see Fig. 1. Since the superpotential $\Delta$, the Darboux
displacement generator $\mathcal{D}_{ab}$ and the factorization
constant $\varepsilon_{ab}$ blow up at $\tau_{ab}=n\K$, $n\in\Z$, we
suppose that  $\tau_{ab}\neq n\K$.

\begin{figure}[h!]\begin{center}
\includegraphics[scale=0.6]{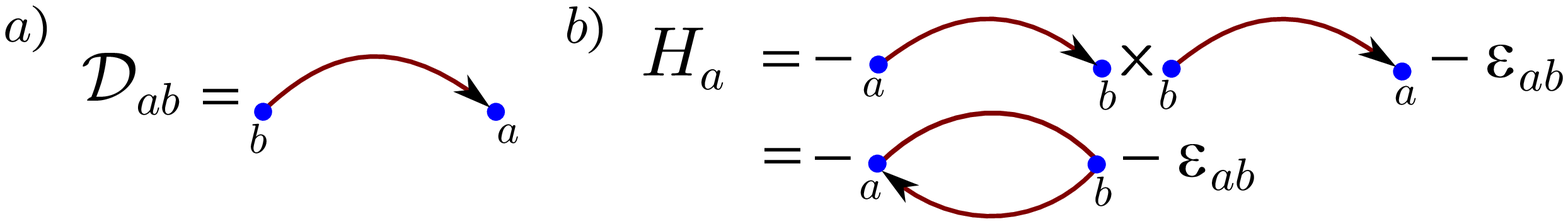}
\caption{ a) The Darboux displacement generator $\mathcal{D}_{ab}$
transforms eigenstates  of $H_b$ into those of the translated system
\textbf{}$H_a$, see (\ref{DabH}). b) The Hamiltonian $H_a$ as a
sequence of the two Darboux displacements (\ref{DDHab}).}
\label{fig1}
\end{center}
\end{figure}

Making use of the relation (\ref{DabH}), one can define the second
order operator
\begin{equation}\label{Bdef}
    \mathcal{B}_{ab/\lambda}=
    \mathcal{D}_{a \lambda}\mathcal{D}^\dagger_{\lambda b}=
    -\mathcal{D}_{a \lambda}\mathcal{D}_{\lambda b}\,,\qquad
    \mathcal{B}^\dagger_{ab/\lambda}=\mathcal{B}_{ba/\lambda}\,,
\end{equation}
where as for $\tau_{ab}$, we assume that
$\tau_{a\lambda},\,\tau_{\lambda b}\neq n\K$. Like the first order
operator $\mathcal{D}_{ab}$, it intertwines the same two systems
$H_a$ and $H_b$,
\begin{equation}\label{BabH}
    \mathcal{B}_{ab/\lambda}H_b=H_a\mathcal{B}_{ab/\lambda},
\end{equation}
via a chain of the two Darboux displacements,
$\mathcal{B}_{ab/\lambda}H_b=
-\mathcal{D}_{a\lambda}\mathcal{D}_{\lambda b}H_b=
-\mathcal{D}_{a\lambda}H_\lambda \mathcal{D}_{\lambda b}= -H_a
\mathcal{D}_{a\lambda}\mathcal{D}_{\lambda b}= H_a
\mathcal{B}_{ab/\lambda}$. In this chain, there appears an
intermediate system $H_\lambda$, which from the viewpoint of our
pair of basic systems $H_a$ and $H_b$ is of a virtual, auxiliary
nature. To stress a virtual nature of the displacement parameter
$\lambda$, we indicate it in a special way (with slash) in notation
for the second order Crum-Darboux intertwining operator
$\mathcal{B}$. {}From (\ref{Bdef}) we  find a relation
\begin{equation}\label{BAAl}
    \mathcal{B}_{aa/\lambda}=\mathcal{D}_{a\lambda}
    \mathcal{D}^\dagger_{a\lambda}=
    H_a+\varepsilon_{a\lambda}\,.
\end{equation}
Hence,  $\mathcal{B}_{ab/\lambda}$ is a kind of a non-Hermitian
generalization of the Lam\'e Hamiltonian operator. In correspondence
with (\ref{BAAl}), the second order intertwiner
$\mathcal{B}_{ab/\lambda}$, unlike $\mathcal{D}_{ab}$, is well
defined (for $\tau_{a\lambda},\, \tau_{\lambda b}\neq n\K$) also in
the case when $\tau_{ab}=n\K$. The virtual parameter on the
right-hand side in (\ref{BAAl}) appears only in the additive term.
We also have
\begin{equation}\label{BBHH}
    \mathcal{B}_{ab/\lambda}\mathcal{B}^\dagger_{ab/\lambda}=
    \mathcal{B}_{ab/\lambda}\mathcal{B}_{ba/\lambda}=
    (H_a+\varepsilon_{a \lambda})(H_a+\varepsilon_{b \lambda})\,.
\end{equation}

Making use of  (\ref{Dgdel2}), we find  that a specific linear
combination of the second, $\mathcal{B}$, and the first order,
$\mathcal{D}$, intertwining operators,
\begin{equation}\label{YBD}
    \mathcal{Y}_{ab}=-\mathcal{B}_{ab/\lambda}-
    g_{ab\lambda}\mathcal{D}_{ab}\,,\qquad
    \mathcal{Y}^\dagger_{ab}=\mathcal{Y}_{ba}\,,
\end{equation}
does not depend on the virtual parameter $\lambda$, where we have
introduced a notation
\begin{equation}\label{gabl}
    g_{ab\lambda}\equiv g(\tau_{ab},\tau_{\lambda a} )=
    \mathfrak{z}(\tau_{ab})+\mathfrak{z}(\tau_{b\lambda})
    +\mathfrak{z}(\tau_{\lambda a})\,
\end{equation}
for a function of displacement parameters, which is  completely
antisymmetric in the indices,
$g_{ab\lambda}=-g_{ba\lambda}=-g_{a\lambda b}$.
 Note the cyclic order of the indices on the r.h.s. of (\ref{gabl}).
 The explicit form of
the intertwining operator
$\mathcal{Y}_{ab}=\mathcal{Y}(x_{ab};\tau_{ab})$,
$\mathcal{Y}_{ab}H_b=H_a\mathcal{Y}_{ab}$, is given by
\begin{equation}\label{Ydef}
    \mathcal{Y}(x;\tau)=
    \frac{d^2}{dx^2}-\Delta(x;\tau)\frac{d}{dx}
    -k^2\sn^2(x+\tau)+{\sn^{-2}2\tau}\,.
\end{equation}
{}From (\ref{YBD}) we get also
\begin{equation}\label{BBdch}
    \mathcal{B}_{ab/\lambda}=\mathcal{B}_{ab/\mu}+
    (g_{ab\mu}-g_{ab\lambda})\mathcal{D}_{ab}\,,
\end{equation}
that corresponds to a change of the virtual displacement parameter,
see Fig. 2. The coefficient in (\ref{BBdch}) before the first order
intertwining operator has a cyclic representation in terms of the
quasi-momentum, $g_{ac\mu}-g_{ac\lambda}=\mathfrak{z}(\tau_{a
\lambda})+ \mathfrak{z}(\tau_{\lambda c})+\mathfrak{z}(\tau_{c
\mu})+ \mathfrak{z}(\tau_{\mu a})$, cf. (\ref{gabl}).

\begin{figure}[h!]\begin{center}
\includegraphics[scale=0.6]{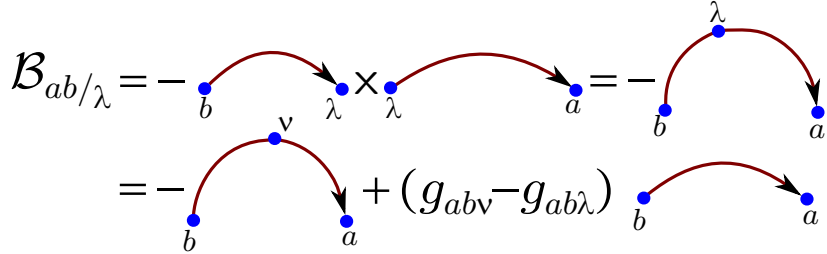}
\caption{ The second order intertwiner as a sequence of the two
Darboux displacements; second line corresponds to a change
$\lambda\rightarrow \nu$ of the virtual parameter
(\ref{BBdch}).}\label{fig2}
\end{center}
\end{figure}

Intertwining operators of the first and the second order allow us to
construct a nontrivial integral for Lam\'e system $H_a$,
\begin{equation}\label{Pa}
    \mathcal{P}_a=\mathcal{D}_{ab}\mathcal{Y}_{ba}+
    \varepsilon_{ab}\mathcal{C}_{ab}=
    \mathcal{Y}_{ab}\mathcal{D}_{ba}-
    \varepsilon_{ab}\mathcal{C}_{ab}\,,
\end{equation}
$[\mathcal{P}_a,H_a]=0$, where
$\mathcal{P}_a=\mathcal{P}(x+\tau_a)$,
$\mathcal{C}_{ab}=\mathcal{C}(\tau_{ab})=-\mathcal{C}_{ba}$,
$\mathcal{C}(\tau)\equiv
g(\tau,\frac{1}{2}\K)=\dn\,2\tau/(\sn\,2\tau\cn\,2\tau)$. Integral
(\ref{Pa}) is nothing else as the Lax operator for one-gap Lam\'e
system (\ref{HLame}), whose explicit form is
\begin{equation}\label{PLax}
    \mathcal{P}(x)=\frac{d^3}{dx^3}+
    (1+k^2-3k^2\sn^2x)\frac{d}{dx}
    -3k^2\sn\, x\cn\, x\dn\, x\,,\qquad
    \mathcal{P}^\dagger=-\mathcal{P}\,.
\end{equation}

Relation (\ref{Pa}) can be presented in the equivalent form
\begin{equation}\label{DDD}
    \mathcal{D}_{ab}\mathcal{D}_{bc}\mathcal{D}_{ca}=
    \mathcal{P}_a-g_{abc}H_a+\varepsilon_{ab}-\xi_{abc}\,,
\end{equation}
where
\begin{equation}\label{xidef}
    \xi_{abc}\equiv\varepsilon_{ab}(g_{abc}-
    \mathcal{C}_{ab})\,.
\end{equation}
Making use of the equivalent representation for the function
(\ref{DDDx}), (\ref{gcs}),
$$
g(\tau;\mu)=\frac{\mathcal{C}(\tau)\varepsilon(\tau)-
\mathcal{C}(\mu)\varepsilon(\mu)}{\varepsilon(\tau)-
\varepsilon(\mu)}\,,
$$
one can check that the three-index object defined in (\ref{xidef})
possesses the same antisymmetry properties as $g_{abc}$,
$\xi_{abc}=-\xi_{bac}=-\xi_{acb}$. We can also write
\begin{equation}\label{PBDg}
    \mathcal{P}_a=-\mathcal{B}_{ab/c}\mathcal{D}_{ba}+
    g_{abc}H_a+\xi_{abc}=
    -\mathcal{D}_{ab}\mathcal{B}_{ba/c}-
    g_{abc}H_a-\xi_{abc}\,,
\end{equation}
see Fig. 3. With the help of (\ref{PBDg}) and relations
\begin{eqnarray}
    \varepsilon_{ab}\varepsilon_{bc}\varepsilon_{c a}
    -\xi_{abc}^2&=&0,\nonumber\\
    \varepsilon_{ab}\varepsilon_{ac}+
    \varepsilon_{ab}\varepsilon_{bc}+
    \varepsilon_{ac}\varepsilon_{bc}
    -2g_{abc}\xi_{abc}&=&k'{}^2,\nonumber\\
    \varepsilon_{ab}+\varepsilon_{ac}+
    \varepsilon_{bc}-g^2_{abc}&=&-(1+k'{}^2)\,,
    \label{eeegx}
\end{eqnarray}
 we find that the Lax integral and the Hamiltonian
 satisfy the Burchnall-Chaundy operator identity
\begin{equation}\label{PP2H}
    -\mathcal{P}^2=P(H)\,,\qquad
    P(H)=H(H-k'{}^2)(H-1)\,,
\end{equation}
where $P(H)$ is a spectral polynomial of the one-gap Lam\'e system
(\ref{HLame}). One can show  that in correspondence with
(\ref{PP2H}), the physical states (\ref{Psipm}) are also the
eigenstates of the Lax operator, $\mathcal{P}\Psi^\alpha_\pm(x)=\mp
i\eta(E(\alpha))\sqrt{P(E(\alpha))}$, where $\eta(E)=-1$ for the
valence and $+1$ for the conduction bands \cite{PAN}. The Lax
integral distinguishes the left- ($\Psi^\alpha_+$) and the right-
($\Psi^\alpha_-$) moving Bloch modes inside these bands, and
annihilates the band-edge states.

\begin{figure}[h!]\begin{center}
\includegraphics[scale=0.6]{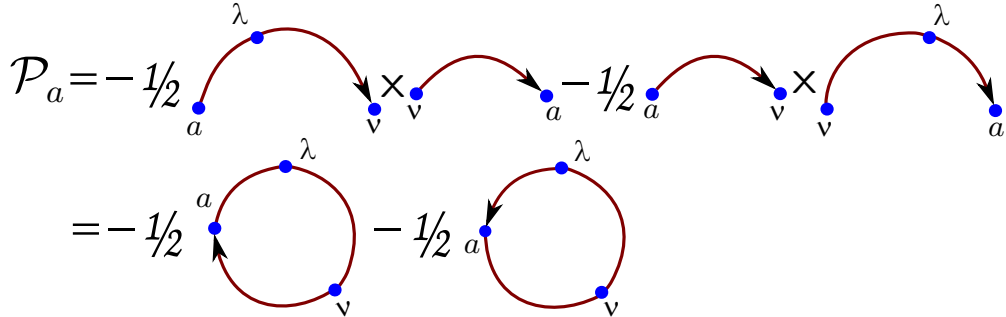}
\caption{ Two representations for the third order (Lax) integral:
$\mathcal{P}_a=-\frac{1}{2}\mathcal{D}_{a\nu}\mathcal{B}_{\nu
a/\lambda} -\frac{1}{2}\mathcal{B}_{a\nu/\lambda}\mathcal{D}_{\nu
a}=-\frac{1}{2}\mathcal{D}_{a\nu}\mathcal{D}_{\nu\lambda}\mathcal{D}_{\lambda
a}-\frac{1}{2}\mathcal{D}_{a\lambda}\mathcal{D}_{\lambda\nu}\mathcal{D}_{\nu
a}$.}\label{fig3}
\end{center}
\end{figure}

{}Proceeding from the definition of the second order operator
$\mathcal{B}_{ab/\lambda}$ as a composition  of the first order
Darboux displacement generators, one can generalize the picture by
treating the intermediate system $H_\lambda$ not just as a virtual
one but on equal grounds as the systems $H_a$ and $H_b$.

The second order intertwining generator can be generalized for the
case of the third order. Making use of the already known lower order
relations and identity (\ref{BBdch}),  one finds
\begin{eqnarray}
    &&\mathcal{D}_{ab}\mathcal{D}_{bc}\mathcal{D}_{cd}=
    -\mathcal{B}_{ac/b}\mathcal{D}_{cd}=-\mathcal{D}_{ab}\mathcal{B}_{bd/c}\nonumber\\
    &&=-(H_a+\varepsilon_{dc})\mathcal{D}_{ad}
    +(g_{acd}-g_{acb})
    \mathcal{B}_{ad/c}
    =
    -(H_a+\varepsilon_{ab})\mathcal{D}_{ad}
    +(g_{bda}-g_{bdc})
    \mathcal{B}_{ad/b}\,.\qquad\quad\label{DDDint}
\end{eqnarray}
The variation in the form of the two last expressions is in the
intermediate indices. Such an ambiguity (freedom) related to the
intermediate indices also appears in the higher order
generalizations of the intertwining relations and integrals (Darboux
loops). If we put $d=a$ in (\ref{DDDint}), the triple product of the
$\mathcal{D}$s reduces to a third order Darboux loop (\ref{DDD}),
while in the two last expressions the first order operator
$\mathcal{D}$ and the coefficient before the second order
intertwining operator are singular. A careful treatment of these
expressions in the limit sense $d\rightarrow a$ reproduces correctly
the r.h.s. of the relation (\ref{DDD}).

  In the next, the fourth order case, we get in the same
way
\begin{eqnarray}
    \mathcal{D}_{ab}\mathcal{D}_{bc}\mathcal{D}_{cd}
    \mathcal{D}_{df}&=&\mathcal{B}_{ac/b}\mathcal{B}_{cf/d}
    =
    (H_a+\varepsilon_{dc}+g_{adc}
    - g_{adf})
    \mathcal{B}_{af/d}\nonumber\\
    &+&(g_{acd}-g_{acb})
    (H_a+\varepsilon_{fd})\mathcal{D}_{af}.\label{DDDDint}
\end{eqnarray}
Taking in the last relation $f=a$, we find that the fourth order
integral for $H_a$ can be presented as follows
\begin{equation}
    \mathcal{D}_{ab}\mathcal{D}_{bc}\mathcal{D}_{cd}
    \mathcal{D}_{da}=(H_a+\varepsilon_{dc})
    (H_a+\varepsilon_{ad})
    +(g_{acb}-
    g_{acd})
    (\mathcal{P}_a+g_{acd}H_a
    +\xi_{acd})\,,
    \label{DDDDinteg}
\end{equation}
i.e. it reduces to a function of $H_a$ and $\mathcal{P}_a$.
Continuing, we find that in the general
 case the closed (loop) sequence of the $n$
Darboux transformations
$\mathcal{D}_{ab_{1}}\mathcal{D}_{b_{1}b_{2}}\ldots
\mathcal{D}_{b_{n-1}a}$ is an integral of motion for the system
$H_a$ of the form
\begin{equation}\label{DintnHP}
    \mathcal{D}_{ab_{1}}\mathcal{D}_{b_{1}b_{2}}\ldots
    \mathcal{D}_{b_{n-1}a}=h_1(H_a)+h_2(H_a)\mathcal{P}_a\,,
\end{equation}
where $h_{1,2}(H_a)$ are certain polynomials of $H_a$. An unclosed
sequence of $n$ Darboux transformations reduces,  analogously, to a
combination of the first and the second order intertwining operators
with coefficients to be some functions of the intertwined
Hamiltonians $H_a$ and  $H_c$,
\begin{equation}
     \mathcal{D}_{ab_{1}}\mathcal{D}_{b_{1}b_{2}}\ldots
    \mathcal{D}_{b_{n-1}c}=
    f_1(H_a)\mathcal{D}_{ac}+
    f_2(H_a)\mathcal{B}_{ac/b_{n-1}}
    =
    \mathcal{D}_{ac}f_1(H_c)+
    \mathcal{B}_{ac/b_{n-1}}f_2(H_c)
    \,.\label{DDHDB}
\end{equation}
Index $b_{n-1}$ in $\mathcal{B}_{ac/b_{n-1}}$ can be changed for any
other intermediate index by employing identity (\ref{BBdch}).

We conclude therefore that the higher order open chains of the
Darboux displacement transformations reduce as differential
operators to linear combinations of the two basic blocks: the first,
$\mathcal{D}$, and the second, $\mathcal{B}$, order Darboux
displacement generators with coefficients to be certain functions of
the intertwined Hamiltonians. In the case of the closed (loop)
chains, they reduce, analogously,   to a linear function of the
third order Lax integral $\mathcal{P}$ with coefficients depending
on the Hamiltonian. No new structures do appear in addition to these
sets of the first and the second order intertwining generators and
the third order Lax integrals, which will play a role of the basic
blocks in the associated supersymmetric construction, to the
discussion of which we pass over in the next section.

\section{Supersymmetry of self-isospectral periodic chains}

In this section we introduce a kind of the $N$-extended system to be
a self-isospectral chain of one-gap Lam\'e systems, and study
general characteristics of the  supersymmetric structure associated
with it.

Consider a chain of $N\geq 2$ one-gap Lam\'e systems which we
describe by a matrix  Hamiltonian
\begin{equation}\label{HchN}
    \mathcal{H}=diag (H_1,\ldots, H_N)\,.
\end{equation}
Here we use the same notations as in (\ref{Dab}), $H_a=H(x+\tau_a)$,
$a=1,\ldots, N$, and assume that the set of the shift parameters
$\tau_a$  is restricted by the condition
$\tau_{ab}=\frac{1}{2}(\tau_b-\tau_a)\neq n\K$ for any pair of
indices $a\neq b$, i.e. we suppose that the arguments of
Hamiltonians of any two subsystems are shifted mutually for any
distance to be different from the real period $n2\K$.

Introduce   a symbol $\he_{ab}$ defined by $\he_{ab}=+1$ for $a<b$,
$\he_{ab}=-1$ for $a>b$, and $\he_{ab}=0$ if $a=b$. We imply that
$(\he_{ab})^{2n+1}=\he_{ab}$, $n\in \Z$,  while $(\he_{ab})^{2n}=+1$
for $a\neq b$ and $(\he_{ab})^{2n}=0$ for $a=b$. We  also introduce
the $N\times N$ matrices:
\begin{equation}\label{Sig1}
    \left(\Si_1^{ab}\right)_{ij}=(\he_{ab})^2(\delta_{i}^{a}\delta_{j}^{b}
    +\delta_{i}^{b}\delta_{j}^{a})\,,\qquad
    \left(\Si_2^{ab}\right)_{ij}=\ii\he_{ab}
    (\delta_{i}^{a}\delta_{j}^{b}-\delta_{i}^{b}\delta_{j}^{a})\,,
\end{equation}
\begin{equation}\label{Sig2}
    \left(\Si_3^{ab}\right)_{ij}=\he_{ab}(\delta_{i}^{a}
    \delta_{j}^{a}-\delta_{i}^{b}\delta_{j}^{b})\,,\qquad
    \left(\bI^{ab}\right)_{ij}=(\he_{ab})^2(\delta_{i}^{a}
    \delta_{j}^{a}+\delta_{i}^{b}\delta_{j}^{b})\,.
\end{equation}
In (\ref{Sig1}) and (\ref{Sig2}) we assume that $a\neq b$, and so,
the first factor in definition of $\Sigma_1^{ab}$ and $\bI^{ab}$ can
be omitted. By definition, all the four matrices are symmetric in
indices $a,b$, $\Si_k^{ab}=\Si_k^{ba}$, $k=1,2,3$,
$\bI^{ab}=\bI^{ba}$. For $N=2$ they reduce to the three Pauli and
the unit matrices, and for $N\geq 2$ satisfy the same algebra
$\Si_i^{ab}\Si_j^{ab}=\delta_{ij}\bI^{ab}+\ii
\epsilon_{ijk}\Si_k^{ab}$, $i,j,k=1,2,3$.

Making use of the intertwining relations from the previous section,
we find that the system (\ref{HchN}) is characterized by the
$N(N-1)$ nontrivial integrals
\begin{equation}\label{Sdef}
    \left(S^{ab}_l\right)_{ij}=(\ii)^{l+1}
    \left(\Si_1^{ab}\right)_{ij}
    (\he_{ij})^l\cD_{ij}\,,
    \qquad l=1,2\,,
\end{equation}
which are the matrix differential operators of the first order, and
by the same number of the integrals of the second order
\begin{equation}\label{Qdef}
    \left(Q^{ab}_{l/\lambda}\right)_{ij}=
    (\delta_{i}^a\delta_{j}^b+\delta_{i}^b\delta_{j}^a)
    (\ii\he_{ij})^{l-1}\cB_{ij/\lambda}\,,
    \qquad l=1,2\,,
\end{equation}
\begin{equation}\label{HSQcom}
    [\cH,S_l^{ab}]=0,\qquad
    [\cH,Q_{l/\lambda}^{ab}]=0\,.
\end{equation}
In definition of the integrals (\ref{Qdef}) we suppose  that the
virtual parameter $\tau_\lambda$ can take independent values for
each pair of indices $a\neq b$, and for any of the two values of the
index $l$; the only restriction, as before, is
$\tau_{a\lambda},\tau_{\lambda b}\neq n\K$. On the other hand,
relation (\ref{BBdch}) means that any second order integral $Q$ with
the changed value of the virtual parameter is a linear combination
of the initial operator $Q$  and of the first order integral $S$.

 In accordance with the introduced notations,
the integrals with  the indices $l=1$ and $l=2$ are related by
\begin{equation}\label{S12S}
    S^{ab}_2=\ii\Si_3^{ab} S^{ab}_1,\qquad
    Q^{ab}_{2/\lambda}=\ii\Si_3^{ab} Q^{ab}_{1/\lambda}\,.
\end{equation}
This is coherent with the fact that  $\bI^{ab}$ and  $\Si_3^{ab}$
are also the integrals of motion for  $\cH$,
$[\cH,\bI^{ab}]=[\cH,\Si_3^{ab}]=0$, and they act, respectively, as
the identity and $\sigma_3=diag(1,-1)$ matrices in the two-term
subsystem specified by the  indices $a\neq b$. Also, the following
relations are valid:
\begin{equation}\label{MW}
    M^{ab}=M^{ba},\qquad
     M^{ab}W^{cd}=0\quad {\rm when\,\,  all\,}\, a,b,c,d\,\,
    {\rm are\,\, distinct}\,,
\end{equation}
 where any of  $M$ and $W$ is $S_l$ or
$Q_{l/\lambda}$.

Matrix
\begin{equation}\label{Gamma}
    \Gamma=diag(1,-1,\ldots, (-1)^{N-1},(-1)^N)\,,\qquad
    \Gamma^2=1\,,
\end{equation}
is an  (zero order) integral of motion for the system (\ref{HchN}),
$[\mathcal{H},\Gamma]=0$, and can be taken as a grading operator. It
identifies the integrals $S_l^{ab}$ and $Q_{l/\lambda}^{ab}$ with
$a-b=2n+1$ as fermionic operators, $\{\Gamma,M^{ab}\}=0$,  and those
with  $a-b=2n$ as  bosonic, $[\Gamma,M^{ab}]=0$. To identify the
superalgebraic structure generated by the integrals of motion, we
compute anti-commutators between  fermionic integrals
(supercharges), and we take commutators between bosonic, and between
bosonic and fermionic integrals. In accordance with (\ref{MW}),
corresponding commutators, $[M^{ab},M^{cd}]$,   and
anti-commutators, $\{M^{ab},M^{cd}\}$, take zero values when all the
indices $a,b,c,d$ are distinct. Nontrivial anti-commutators for
fermionic supercharges are
\begin{equation}\label{A1}
    \{S^{ab}_l,S^{ab}_m\}=
    2\delta_{lm}\bI^{ab}(\cH+\varepsilon_{ab})\,,
\end{equation}
\begin{equation}\label{A2}
    \{S^{ab}_l,S^{bc}_m\}=(\he_{ab})^l(\he_{bc})^m
    \left(\delta_{lm}(-1)^{l} Q^{ac}_{1/b}+
    (1-\delta_{lm})\he_{ac}Q^{ac}_{2/b}\right)\,,
\end{equation}
\begin{equation}\label{A3}
    \{Q^{ab}_{l/\lambda},Q^{ab}_{m/\lambda}\}=
    2\delta_{lm}\bI^{ab}(\cH+
    \varepsilon_{a\lambda})(\cH+\varepsilon_{b\lambda})\,,
\end{equation}
\begin{equation}\label{A4}
    \{Q^{ab}_{l/\lambda},Q^{bc}_{m/\lambda}\}=
    (\he_{ab})^{l-1}(\he_{bc})^{m-1}\left(\delta_{lm}(-1)^{l-1}
    Q^{ac}_{1/\lambda}
    +(1-\delta_{lm})\he_{ac}Q^{ac}_{2/\lambda}
    \right)(\cH+\varepsilon_{b\lambda})\,,
\end{equation}
\begin{equation}
    \label{L1}
    \{Q^{ab}_{l/\lambda},S^{ab}_{m}\}=
    2(\he_{ab})^{l-1}(\he_{ba})^{m}\bI^{ab}\left(\delta_{lm}(-1)^{l}
    (g_{ab\lambda}\cH +\xi_{ab\lambda})-
    (1-\delta_{lm})L\right)\,,
\end{equation}
\begin{equation}\label{A5}
    \{Q^{ab}_{l/\lambda},S^{bc}_{m}\}=
    (\he_{ab})^{l-1}(\he_{bc})^{m}\left(\delta_{lm}(-1)^{l+1}
    \he_{ac}S^{ac}_{1}
    +(1-\delta_{lm})S^{ac}_{2}\right)(\cH+\varepsilon_{bc})\,.
\end{equation}
The nontrivial commutators are given by
 \begin{equation}\label{AA1}
     [ S^{ab}_l,S^{ab}_m ]=-2\ii(\he_{ab})^{l+1}
     (\he_{ba})^m(1-\delta_{lm})\Si_3^{ab}(\cH+\varepsilon_{ab})\,,
\end{equation}
\begin{equation}\label{AA2}
     [ S^{ab}_l,S^{bc}_m ]=\ii(\he_{ab})^l(\he_{bc})^m
     \left(\delta_{lm}(-1)^{l+1}\he_{ac} Q^{ac}_{2/b}-
     (1-\delta_{lm})Q^{ac}_{1/b}\right)\,,
\end{equation}
\begin{equation}\label{AA3}
     [Q^{ab}_{l/\lambda},Q^{ab}_{m/\lambda}]=
     -2\ii(\he_{ab})^{l}(\he_{ba})^{m-1}
     (1-\delta_{lm})\Si_3^{ab}(\cH+
     \varepsilon_{a\lambda})(\cH+\varepsilon_{b\lambda})\,,
\end{equation}
\begin{equation}\label{AA4}
     [Q^{ab}_{l/\lambda},Q^{bc}_{m/\lambda}]=
     \ii(\he_{ab})^{l-1}(\he_{bc})^{m-1}
     \left(\delta_{lm}(-1)^l\he_{ac} Q^{ac}_{2/\lambda}+
     (1-\delta_{lm})Q^{ac}_{1/\lambda}\right)(\cH+\varepsilon_{b\lambda})\,,
\end{equation}
\begin{equation}\label{L2}
     [Q^{ab}_{l/\lambda},S^{ab}_{m}]=
     2\ii (\he_{ab})^{l}(\te_{ba})^{m}\Si_3^{ab}\left((-1)^{l+1}
     \delta_{lm}L+(1-\delta_{lm})(g_{ab\lambda}\cH +
     \xi_{ab\lambda})\right)\,,
\end{equation}
\begin{equation}\label{AA5}
     [Q^{ab}_{l/\lambda},S^{bc}_{m}]=
     \ii(\he_{ab})^{l-1}(\he_{bc})^{m}
     \left(\delta_{lm}(-1)^l S^{ac}_{2}+(1-
     \delta_{lm})\he_{ac}S^{ac}_{1}\right)
(\cH+\varepsilon_{bc})\,.
\end{equation}
In correspondence with the comment on the change of the virtual
index we made above, without loss of generality
 we put the same value for it
in both second order integrals in each of the corresponding
(anti)-commutators
 in (\ref{A3}), (\ref{A4}),
(\ref{AA3}) and (\ref{AA4}).

In (\ref{L1}) and  (\ref{L2}) there appears a nontrivial matrix
operator
\begin{equation}
    L=-\ii\,
    diag(\mathcal{P}_1,\mathcal{P}_2,\cdots,
    \mathcal{P}_{N})\,,
\end{equation}
composed from the third order Lax operators
$\mathcal{P}_{a}=\mathcal{P}(x+\tau_a)=
-\half(\cD_{ab}\cB_{ba/\lambda}+\cB_{ab/\lambda}\cD_{ba})$, see Eq.
(\ref{Pa}). The bosonic integral $L$  is a central element of the
superalgebra with the grading operator $\Gamma$,
\begin{equation}
    \label{comL1}
    [L,\cH]=[\Gamma, L]=0,\qquad
    [L,S_l^{ab}]=[L,Q_{l/\lambda}^{ab}]=0\,.
\end{equation}
For the self-isospectral chain system (\ref{HchN}),  we have  $N$
obvious (bosonic for (\ref{Gamma})) third order integrals $L^a=-\ii
\,diag (0,\ldots,0, \mathcal{P}_a,0,\ldots, 0)$,  the sum of which
corresponds to the central element $L$. To write the commutation
relations for them in the general case of the $N$-terms chain, it is
convenient to define the linear combinations of $L^a$,
$L^{ab}_1=\bI^{ab}L$ and $L^{ab}_2=\Si_3^{ab}L$, remembering that
for $N>2$ not all them are linearly independent. All these third
order integrals commute between themselves, while their nontrivial
commutators with the first and the second order integrals are
\begin{equation}\label{B1}
    [L^{ab}_l,S^{ab}_m]=
    2\ii^m(\he_{ab})^{m+1}\delta_{l2}(\Si_3^{ab})^{m+1}[LS]^{ab}\,,
\end{equation}
\begin{equation}\label{B2}
    [L^{ab}_l,S^{bc}_m]=
    \ii^m(\he_{ba})^{l-1}\he_{bc}(\Si_3^{bc})^{m+1}[LS]^{bc}\,,
\end{equation}
\begin{equation}\label{B3}
    [L^{ab}_l,Q^{ab}_{m/\lambda}]=
    -2\ii^m\delta_{l2}(\Si_3^{ab})^{m}[LQ]^{ab}_{/\lambda}\,,
\end{equation}
\begin{equation}\label{B4}
    [L^{ab}_l,Q^{bc}_{m/\lambda}]=
    -\ii^{m+1}(\he_{ba})^{l-1}\he_{bc}(\Si_3^{bc})^m
    [LQ]^{bc}_{/\lambda}\,,
\end{equation}
where $l=1,2$, and we denote
\begin{equation}\label{LabS}
    [LS]^{ab}\equiv Q^{ab}_{1/\lambda}(\cH+\varepsilon_{ab})
    -\he_{ab}S^{ab}_1(g_{ab\lambda}\cH +\xi_{ab\lambda})\,,
\end{equation}
\begin{equation}\label{LabQ}
    [LQ]^{ab}_{/\lambda}\equiv
    \he_{ab}Q^{ab}_{2/\lambda}(g_{ab\lambda}\cH+\xi_{ab\lambda})+S^{ab}_2\big(\cH(\cH
    + 2k'^2\varepsilon_{ab}^{-1}-\varepsilon_{a\lambda}-\varepsilon_{b\lambda}
    )+\varepsilon_{a\lambda}\varepsilon_{b\lambda }\big)\,.
\end{equation}
Though on the right hand side of (\ref{LabS}), there appears
explicitly a virtual parameter $\lambda$, the complete combination
of the integrals there does not depend on $\lambda$. This can be
checked by making use of definitions (\ref{Sdef}), (\ref{Qdef}) and
Eqs. (\ref{YBD}) and (\ref{xidef}).

We see that the set of the first, the second and the third order
integrals of motion, which are Hermitian matrix operators,  generate
a kind of nonlinear superalgebra. A nonlinearity is related to the
fact that some of the (anti)-commutators of these integrals are
quadratic in the Hamiltonian $\mathcal{H}$, or include $\mathcal{H}$
or $\mathcal{H}^2$ as a multiplier at other integrals. The results
on the general form of intertwining relations from the previous
section show that no new independent integrals do appear in addition
to those we already found. The interesting property of this
supersymmetric structure is also that the set of the second order
integrals $Q^{ab}_{l/\lambda}$ taken \emph{with the same value for
the virtual parameter $\lambda$} (i.e. with the same shift parameter
$\tau_\lambda$) together with the Hamiltonian $\mathcal{H}$ form a
closed nonlinear sub-superalgebra, see Eqs. (\ref{A3}), \ref{A4}),
(\ref{AA3}) and (\ref{AA4}). The second order integrals can  be
reduced to a `standard' form with a prescribed, (any) fixed value of
the virtual parameter by means of relation (\ref{BBdch}).

\section{Alternative choices for  the $\Z_2$-grading operator}

The choice (\ref{Gamma}) for the grading operator is not unique. The
permutation of diagonal elements in (\ref{Gamma}), or multiplication
of some of them by $-1$ changes the identification of integrals as
fermionic and bosonic ones, and, as a consequence, some commutation
relations will be changed for anti-commutation relations and vice
versa. This does not change, however, the bosonic nature of the
diagonal matrix integrals $L^{ab}_l$, and a global  conclusion on a
nonlinear nature of superalgebra.

There are alternative choices for the grading operator which involve
reflections in the coordinate $x$ and in the shift parameters. They
provide some new features for the superalgebraic structure. Let us
discuss some of such alternatives. Consider the reflection in $x$
(parity) operator $\mathcal{R}$, $\mathcal{R}x=-x\mathcal{R}$,
$\mathcal{R}\tau_i=\tau_i\mathcal{R}$, $\mathcal{R}^2=1$, and the
operator $\mathcal{T}$ that  reflects any of  the shift parameters,
including the virtual ones, $\mathcal{T}\tau_i=-\tau_i\mathcal{T}$,
$\mathcal{T}x=x\mathcal{T}$, $\mathcal{T}^2=1$, and commutes with
$\mathcal{R}$. The product of these two operators is a nontrivial,
nonlocal integral of motion for our chain system,
$[\mathcal{RT},\mathcal{H}]=0$, and its square equals $1$. So, it
can be identified as another sort of the grading operator,
\begin{equation}\label{RTG}
    \hat{\Gamma}=\mathcal{RT}\,.
\end{equation}
The operator $\hat{\Gamma}$ anti-commutes with all the first,
$S^{ab}_l$, and the third, $L^{ab}_l$,  order integrals, and
commutes with the second order integrals $Q^{ab}_{l/\lambda}$. In
this case $S^{ab}_l$ and  $L^{ab}_l$ are the fermionic operators,
while $Q^{ab}_{l/\lambda}$ are the bosonic ones. To compute the
superalgebraic structure for such a choice of the grading operator,
we have to use coherently with the described identification of
bosonic and fermionic generators  the corresponding
(anti)-commutators from the previous section, which should be
supplied with the nontrivial anti-commutation relations that involve
the third order integrals,
\begin{equation}\label{BB1}
    \{L^{ab}_l,S^{ab}_m\}=2\ii^m(\he_{ab})^{m+1}
    \delta_{l1}(\Si_3^{ab})^{m
    }[LS]^{ab}\,,
\end{equation}
\begin{equation}\label{BB2}
    \{L^{ab}_l,S^{bc}_m\}=\ii^m
    (\he_{ba})^{l-1}(\Si_3^{bc})^{m}[LS]^{bc}\,,
\end{equation}
\begin{equation}\label{BB3}
    \{L^{ab}_l,Q^{ab}_{m/\lambda}\}=
    -2\ii^m\delta_{l1}(\Si_3^{ab})^{m+1}[LQ]^{ab}_{/\lambda}\,,
\end{equation}
\begin{equation}\label{BB4}
    \{L^{ab}_l,Q^{bc}_{m/\lambda}\}=
    -\ii^{m+1}(\he_{ba})^{l-1}
    (\Si_3^{bc})^{m+1}[LQ]^{bc}_{/\lambda}\,,
\end{equation}
\begin{equation}\label{BB5}
    \{L^{ab}_l,L^{ab}_{m}\}=
    2(\Si_3^{ab})^{l+m}P(\mathcal{H})\,,
\end{equation}
\begin{equation}\label{BB6}
    \{L^{ab}_l,L^{bc}_{m}\}=
    (\bI^{ab}-\hat{e}_{ab}\Sigma_3^{ab})(\he_{ba})^{l-1}
    (\he_{bc})^{m-1}P(\mathcal{H})\,.
\end{equation}
Under such a choice of the grading operator, the spectral polynomial
of the chain, $P(\mathcal{H})=
\mathcal{H}(\mathcal{H}-k'{}^2)(\mathcal{H}-1)$, appears explicitly
in the superalgebraic structure.
 This choice, therefore, is coherent with the structure of a hidden,
bosonized supersymmetry  (\ref{hidnsusy}) \cite{MP1} that is present
in each of the chain subsystems $H_a$ \cite{CP1}. The relation
(\ref{BB5}) particularly reveals the property of the Lax operators
that is essential for physical applications: each third order
differential operator $L_l$  here is an annihilator of the three
band edge states in the spectrum of the corresponding chain Lam\'e
subsystem $H_l$, see refs. \cite{CJNP,PAN,CJPJA} for the further
details. In the case of the choice of the grading operator discussed
in the previous section, this peculiarity of the Lax operators does
not show up in the superalgebraic relations.

The product of the two operators, (\ref{Gamma}) and (\ref{RTG}), can
also  be chosen as a grading operator. Yet other possibilities are
associated with the introduction of the operators of the
permutations of the displacement parameters, $T^{ab}=T^{ba}$,
defined by $T^{ab}\tau_b=\tau_a T^{ab}$,
$T^{ab}\tau_c=\tau_cT^{ab}$, $T^{ab}x=xT^{ab}$. Combining such
operators with the matrix structures $\Sigma^{ab}_{1,2,3}$, and
reflection operators $\mathcal{R}$ and $\mathcal{T}$, one can
construct more integrals which can be taken as the grading
operators. This does not add something essentially new to the
structures we already observed, and we do not discuss these other
possibilities here.

In the last section we present some further arguments in favor of
necessity to consider alternative choices for the grading operator
(alongside with the choice discussed in the previous section) in the
context of possible physical applications.

\section{Supersymmetric structure of  the $N=3$ chain}

In the case of the two-term chain ($N=2$), we have $a,b=1,2$. A
complete set of independent integrals is formed by  the two
integrals of the first order, $S^{12}_l$,  $l=1,2,$ the two
integrals of the second order, $Q^{12}_{l/\lambda}$, and by the two
integrals of the third order, $L^{12}_1=L$ and $L^{12}_2$. In the
list of the (anti)-commutation relations of the integrals there do
not appear (anti)-commutators which involve the generators with
three different indices $a,b,c$. For the discussion of the case
$N=2$ we refer to \cite{PAN}. Here we consider in more detail the
next case $N=3$ to illustrate explicit matrix form of the involved
structures.

The $N=3$ Hamiltonian is
\begin{equation}
    \cH=\begin{pmatrix} H_1 & 0 & 0 \\
    0 & H_2 & 0 \\0  & 0  & H_3
\end{pmatrix}.
\end{equation}
The system has six trivial matrix integrals of motion,
\begin{equation}
    \bI^{12}=
    \begin{pmatrix} 1 & 0 & 0 \\ 0 & 1 & 0 \\0  & 0  & 0
    \end{pmatrix},\quad \bI^{13}=
    \begin{pmatrix} 1 & 0 & 0 \\0 & 0 & 0
    \\0  & 0  & 1 \end{pmatrix},\quad \bI^{23}=
    \begin{pmatrix} 0 & 0& 0
    \\0 & 1 & 0 \\0  & 0  & 1
\end{pmatrix},
\end{equation}
\begin{equation}
    \Si_3^{12}=
    \begin{pmatrix} 1 & 0 & 0 \\ 0 & -1 & 0 \\0  & 0  & 0
    \end{pmatrix},\quad \Si_3^{13}=
    \begin{pmatrix} 1 & 0 & 0 \\0 & 0 & 0
    \\0  & 0  & -1 \end{pmatrix},\quad \Si_3^{23}=
    \begin{pmatrix} 0 & 0
    & 0 \\0 & 1 & 0 \\0  & 0  & -1
    \end{pmatrix},
\end{equation}
which appear explicitly in the superalgebraic (anti)-commutation
relations. This set contains only three linearly independent
matrices.  Three nontrivial integrals to be the first order
differential operators are
\begin{equation}
     S^{12}_1=\begin{pmatrix} 0 & -\cD_{12} & 0
    \\\cD_{21} & 0 & 0 \\0  & 0  & 0
    \end{pmatrix},\quad S^{13}_1=\begin{pmatrix} 0 & 0 & -\cD_{13} \\0 &
    0 & 0 \\\cD_{31}  & 0  & 0 \end{pmatrix},\quad
    S^{23}_1=\begin{pmatrix} 0 & 0 & 0 \\0 & 0 & -\cD_{23} \\0  &
    \cD_{32}  & 0 \end{pmatrix},
\end{equation}
and the other  three are $S^{ab}_2=i\Si_3^{ab}S^{ab}_1$. Six second
order integrals of motion are given by
\begin{equation}
    Q^{12}_{1/\lambda}=\begin{pmatrix} 0 & \cB_{12/\lambda} & 0
    \\\cB_{21/\lambda} & 0 & 0 \\0  & 0  & 0
    \end{pmatrix},
    \quad Q^{13}_{1/\lambda}=\begin{pmatrix} 0 & 0 &
    \cB_{13/\lambda} \\0 & 0 & 0
    \\\cB_{31/\lambda}  & 0  & 0 \end{pmatrix},\quad
    Q^{23}_{1/\lambda}=\begin{pmatrix} 0 & 0 & 0 \\0 & 0 &
    \cB_{23/\lambda} \\0
    & \cB_{32/\lambda}  & 0 \end{pmatrix},
\end{equation}
and  $Q^{ab}_{2/\lambda}=i\Si_3^{ab}Q^{ab}_{1/\lambda}$. Finally,
the set of the three linearly independent third order integrals is
\begin{equation}
    L^{12}=\begin{pmatrix} -\ii \mathcal{P}_1 & 0 & 0
    \\0 &-\ii \mathcal{P}_2 & 0 \\0 & 0
    & 0 \end{pmatrix},\quad L^{13}=
    \begin{pmatrix}-\ii
    \mathcal{P}_1 & 0 & 0 \\0
    & 0 & 0 \\0  & 0  &-\ii
    \mathcal{P}_3
    \end{pmatrix},\quad
    L^{23}=\begin{pmatrix} 0 & 0 & 0 \\0 &-\ii
    \mathcal{P}_2 & 0 \\0  & 0 &-\ii
    \mathcal{P}_3 \end{pmatrix}.
\end{equation}
Their linear combination corresponds to the integral $L$,
$L=\frac{1}{2}(L^{12}+L^{13}+L^{23})$.

The grading operator can be chosen in one of the
forms\footnote{There are other possibilities, not reducible to the
change of indices and multiplication of the matrix elements by $-1$
(see the remark at the end of the previous section), which we do not
consider here.}
\begin{equation}\label{3G}
    \Gamma=\begin{pmatrix} 1 & 0 & 0 \\ 0 & -1 & 0 \\0  & 0  &
    1\end{pmatrix},\quad \hat\Gamma=
    \begin{pmatrix} \mathcal{RT} & 0 & 0 \\0 &
    \mathcal{RT}
    & 0 \\0  & 0  & \mathcal{RT} \end{pmatrix},
    \quad \hat\Gamma_1=\Gamma\hat{\Gamma}=\begin{pmatrix}
    \mathcal{RT} & 0 & 0 \\0 & -\mathcal{RT} & 0 \\0  & 0  &
    \mathcal{RT}
    \end{pmatrix}.
\end{equation}
When $\Gamma$ is chosen as the grading operator, we have eight
nontrivial fermionic integrals of motion $S^{12}_l,$ $S^{23}_l,$
$Q^{12}_{l/\lambda},$ $Q^{23}_{l/\lambda}$, and seven linear
independent bosonic integrals $S^{13}_l,$ $Q^{13}_{l/\lambda}$, and
$L^{12}$, $L^{13}$ and $L^{23}$. In the case of the choice of
$\hat\Gamma$ as the grading operator, we have nine fermionic
integrals   $S^{ab}_l$ and  $L^{ab}$. The second order integrals
$Q^{ab}_{l/\lambda}$ constitute the set of six bosonic integrals of
motion. Finally, for $\hat\Gamma_1$, we have six bosonic integrals,
$S^{12}_l$, $S^{23}_l$ and  $Q^{13}_{l/\lambda}$, and nine fermionic
integrals, $S^{13}_{l}$, $Q^{12}_{l/\lambda}$, $Q^{23}_{l/\lambda}$,
and $L^{ab}$. We see that the complete set of local nontrivial
integrals of motion separates into bosonic and fermionic generators
in dependence on the choice of the grading operator.

If we start from the first order integrals $S^{12}_l$, $S^{23}_l$
and $S^{13}_l$, their corresponding (anti)-commutation relations
(\ref{A2}) and/or (\ref{AA2}) (that depends on the choice of the
grading operator) generate, unlike the $N=2$ case, all the six
second order integrals $Q^{ab}_{l/\lambda}$. The virtual parameter
$\tau_\lambda$ here as well as for $N>3$ can be identified with the
shift parameter $\tau_b$ of one of the corresponding subsystems.
Each time, however,  the intermediate index $\lambda$ in
$Q^{ab}_{l/\lambda}$ can be changed by employing relation
(\ref{BBdch}). Anyway, for $N\geq 2$ the second order integrals
$Q^{ab}_{l/\lambda}$ are generated also via the (anti)-commutators
of $S^{ab}_l$ with the third order integrals, see Eqs. (\ref{B1}),
(\ref{B2}), (\ref{LabS}), (\ref{BB1}) and (\ref{BB2}).

\section{Self-isospectral soliton chains}

Consider now  the infinite period limit which produces
self-isospectral non-periodic chains. It is obtained by putting
$k\rightarrow 1$, when, as we noted, $\K\rightarrow \infty$,
$2\ii\K'\rightarrow \ii\pi$, and one-gap Lam\'e Hamiltonian
(\ref{HLame}) transforms into that of reflectionless P\"oschl-Teller
system
\begin{equation}\label{HlimPT}
    H^{PT}(x)=-\frac{d^2}{dx^2}-\frac{2}{\cosh^2x} +1\,.
\end{equation}
In this limit the valence band $0\leq E \leq k'{}^2$ of the Lam\'e
system shrinks into the level $E=0$ of the unique bound state of the
system (\ref{HlimPT})  described by the wave function ${\rm sech}\,
x$. To get a self-isospectral supersymmetric chain of reflectionless
P\"oschl-Teller systems, we have to introduce also some restrictions
on the displacement parameters. Namely, it is necessary to require
that all the $\tau_a$ which appear in the arguments of the chain
Hamiltonians, $H_a=H(x+\tau_a)$, should \emph{not} go to infinity
when $k\rightarrow 1$.  By this condition we prohibit that in the
chain we get in the limit, there could appear free particle systems.

In such a limit, for $x$-independent structures we get
\begin{equation}\label{lim1}
    \mathfrak{z}(\tau)\rightarrow\coth 2\tau\,,\qquad
    \mathcal{C}_{ab}\rightarrow\coth 2\tau_{ab}\,,\qquad
    g_{ab\lambda}\rightarrow \coth 2\tau_{ab} + \coth 2\tau_{b\lambda}
    +\coth 2\tau_{\lambda a}\,,
\end{equation}
\begin{equation}\label{lim2}
    \varepsilon_{ab}\rightarrow \csch^2 2\tau_{ab}\,,\qquad
    \xi_{ab\lambda}\rightarrow -(\coth 2\tau_{a\lambda}+\coth
2\tau_{\lambda b})\csch^2 2\tau_{ab}\,.
\end{equation}
For the superpotential (\ref{Deldef}) we have
$\Delta(x;\tau)\rightarrow \Delta^{PT}(x;\tau)=
\tanh(x-\tau)-\tanh(x+\tau)+\coth 2\tau$. Denoting  the limit of
$\Delta(x_{ab};{\tau_{ab}})$ by $\Delta_{ab}^{PT}$, we find
\begin{eqnarray}
    \Delta_{ab}^{PT}=-\Delta_{ba}^{PT}&=& \tanh(x+\tau_{a})
    -\tanh(x+\tau_{b})+\coth (\tau_{b}-\tau_a)\nonumber\\
    &=& \tanh 2\tau_{ab}
    \tanh(x+\tau_a)\tanh(x+\tau_b)+{2}{\csch 4\tau_{ab}}\,.
    \label{DelPT}
\end{eqnarray}
For the first order intertwining operator we get
\begin{equation}\label{DlimX}
    \mathcal{D}_{ab}\rightarrow
    X_{ab}=\frac{d}{dx}-\Delta_{ab}^{PT}\,,\qquad
    X_{ab}^\dagger=-X_{ba}\,,
\end{equation}
and (\ref{DDHab}) transforms then into
\begin{equation}\label{XXH}
    -X_{ab}X_{ba}=H^{PT}_a+\csch^2 2\tau_{ab}\,,
\end{equation}
where $H^{PT}_a=H^{PT}(x+\tau_a)$. The limit of the second order
intertwining operator (\ref{Ydef}) can be written  in the form that
includes in its structure the first order operator
$A_\tau=\frac{d}{dx}-\tanh(x+\tau)$,
\begin{eqnarray}
    \mathcal{Y}_{ab}&\rightarrow&
    \frac{d^2}{dx^2}-\Delta_{ab}^{PT}\frac{d}{dx}
    -\tanh^2(x+\tau_a)+\coth^22\tau_{ab}
      \nonumber\\
    &=&
    -A_{a}A^\dagger_{b}-\coth 2\tau_{ab}X_{ab}\,,
    \label{YlimPT}
\end{eqnarray}
where $A_{a}\equiv A_{\tau_a}$. The first term in the last
expression in (\ref{YlimPT}), unlike the $X_{ab}$ and the
$x$-independent multiplier $\coth 2\tau_{ab}$ in the second term, is
well defined when $\tau_b=\tau_a\Rightarrow \tau_{ab}=0$. This is so
because the operator $A_\tau$, unlike the limit of the operator
$\mathcal{D}(x;\tau)$, is regular  for $\tau=0$.

Making use of Eqs.  (\ref{YBD}) and (\ref{lim1}), for the limit of
the second order intertwining operator $\mathcal{B}$ we get
\begin{equation}\label{BlimPT}
    \mathcal{B}_{ab/\lambda}\rightarrow
    -X_{a\lambda}X_{\lambda b}=
    A_{a}A^\dagger_{b}
    +(\coth 2\tau_{a\lambda}+
    \coth 2\tau_{\lambda b})
    X_{ab}\equiv B_{ab/\lambda}\,,
\end{equation}
and find that the limit of the third order integral can be presented
in terms of the introduced first order operator $A_\tau$,
\begin{equation}\label{PlimPT}
    \mathcal{P}_a\rightarrow
    -Z_{a}\,,\quad
    {\rm where}\quad
    Z_a=A_{a} \frac{d}{dx}
    A_{a}^\dagger\,.
\end{equation}
{}Since the first order operator $X_{ab}$ as well as the second
order operators (\ref{YlimPT}) and (\ref{BlimPT}) intertwine the
Hamiltonians $H^{PT}_b$ and $H^{PT}_a$, the operator \cite{PNPRD}
\begin{equation}\label{YAA}
    Y_{ab}=A_aA^\dagger_b
\end{equation}
is also the intertwining operator, $Y_{ab}H_b^{PT}=H_a^{PT}Y_{ab}$.
Operator (\ref{YAA}) corresponds to the infinite limit of the
virtual displacement parameter, $\tau_\lambda\rightarrow\infty$ (or,
$\tau_\lambda\rightarrow-\infty$), applied to $B_{ab/\lambda}$, see
Fig. 4.

The intertwiner  (\ref{YAA}), unlike (\ref{YlimPT}) and
(\ref{BlimPT}), is regular for $\tau_{ab}=0$ ($\tau_a=\tau_b$), when
it reduces just to $H_a^{PT}$,
\begin{equation}\label{YaaH}
    Y_{aa}=A_aA^\dagger_a=H_a^{PT}.
\end{equation}
Another product of the same operators produces (for any value of the
parameter $\tau_a$)  the free particle Hamiltonian shifted for an
additive constant,
\begin{equation}\label{H0AA}
    A^\dagger_a A_a=
    -\frac{d^2}{dx^2}+1\equiv H_0\,.
\end{equation}
{}In accordance with (\ref{YaaH}) and (\ref{H0AA}), the first order
operators $A_a$ and $A^\dagger_a$ intertwine the P\"oschl-Teller
system with a free particle, $A^\dagger_a H_a^{PT}=H_0 A^\dagger_a$,
$H_a^{PT}A_a=A_a H_0$.   From (\ref{DelPT}) we find that if
$\tau_b\rightarrow \infty$ while $\tau_a$ is kept to be finite,
$\Delta^{PT}_{ab}$ reduces to $\tanh (x+\tau_a)$. In such a limit,
$H_b^{PT}$ reduces to $H_0$, $H_\infty^{PT}=H_0$, $X_{ab}$
transforms into $A_a$, while the second order operators
(\ref{YlimPT}), (\ref{BlimPT}) and (\ref{YAA}) transform into
$A_a\frac{d}{dx}$ and linear combinations of this operator and the
first order operator $A_a$. The second order operator we have gotten
intertwines $H_a^{PT}$ with $H_0$,
$A_a\frac{d}{dx}H_0=H_{a}^{PT}A_a\frac{d}{dx}$, but this relation
produces nothing new since it is a consequence of the conservation
of $\frac{d}{dx}$ for a free particle system $H_0$,
$\frac{d}{dx}H_0=H_0\frac{d}{dx}$, and of the already known relation
$H_a^{PT}A_a=A_a H_0$. On the other hand, the product of this
operator with the intertwiner $A^\dagger_a$, that acts in another
direction between $H_0$ and $H^{PT}_a$, shows that the Lax operator
of the P\"oschl-Teller system, $Z_a$, is nothing else as the
Darboux-dressed free particle momentum  \cite{CJP1}. {}From
(\ref{BlimPT}) one can get a relation which involves the first order
intertwiners
 $A_a$ and  $X_{ab}$, and the free particle momentum.
 Taking the limit $\tau_b\rightarrow\infty$
in both representations for $B_{ab/\lambda}$, we get the relation $
A_a\left(\frac{d}{dx}-\coth
    2\tau_{a\lambda}\right)=X_{a\lambda}A_\lambda\,,$
and also  its conjugate, $   A_\lambda^\dagger
    X_{\lambda a}= \left(\frac{d}{dx}+\coth
    2\tau_{a\lambda}\right)A_a^\dagger.$

\begin{figure}[h!]\begin{center}
\includegraphics[scale=0.59]{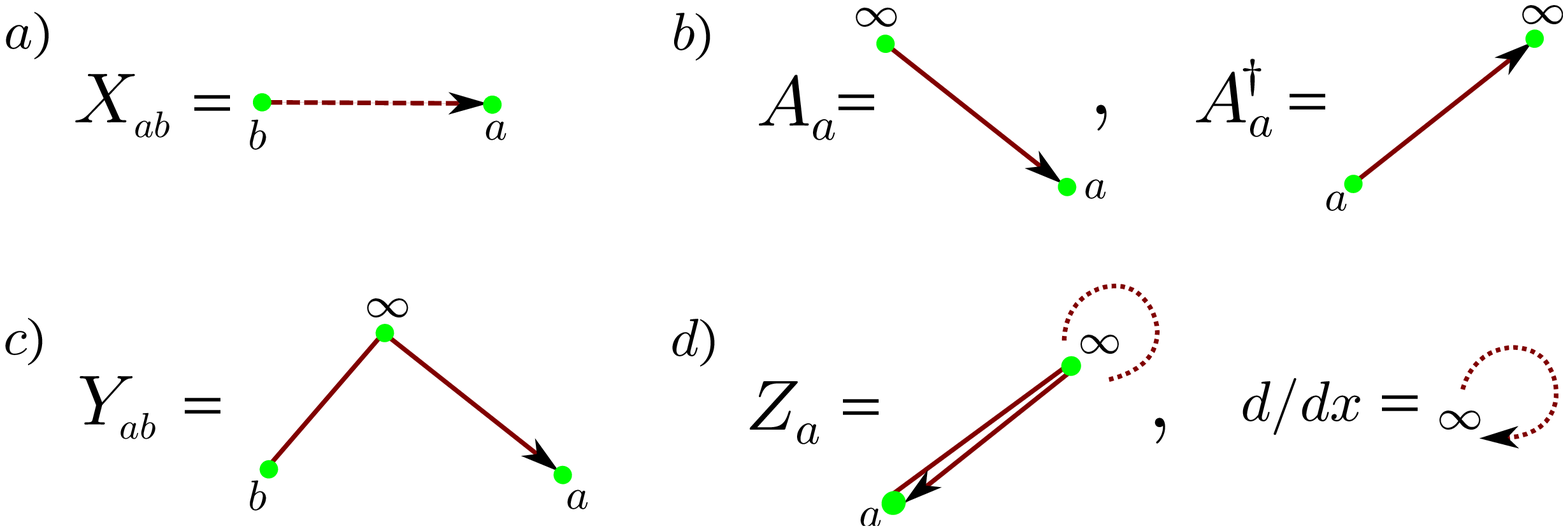}
\caption{The first order Darboux displacement generator $X_{ab}$
translates the eigenfunctions of $H^{PT}_b$ into those of
$H^{PT}_a$.  The second order intertwiner $Y_{ab}$ makes the same
via the virtual free particle system that is a P\"oschl-Teller
system translated to infinity. Lax integral $Z_a$ is presented as a
Darboux dressed form (\ref{PlimPT}) of the free particle integral
$d/dx$.}\label{fig4}
\end{center}
\end{figure}

The described limit applied to the  chain (\ref{HchN}) gives a
non-periodic self-isospectral chain described by the Hamiltonian
\begin{equation}\label{HchN1}
    \mathcal{H}^{PT}=
    diag (H_1^{PT},\ldots, H_N^{PT})\,.
\end{equation}
The  nontrivial integrals of such a system are given by the infinite
period limit applied to the integrals $S^{ab}_l$,
$Q^{ab}_{l/\lambda}$ and $L^{ab}_l$ by employing relations
(\ref{DlimX}), (\ref{BlimPT}) and (\ref{PlimPT}). As in the periodic
case, the set of the corresponding third order integrals
$L^{PTab}_l$ contains only $N$ independent integrals. Instead of
$Q^{PTab}_{l/\lambda}$, one can work with the second order matrix
integrals $\mathcal{Q}^{PTab}_{l}=
\lim_{\tau_\lambda\rightarrow\infty}Q^{PTab}_{l/\lambda}$, which are
obtained by the change of ${B}_{ab/\lambda}$ for $Y_{ab}$. The
corresponding superalgebra generated by these integrals, as in the
periodic case, depends on the choice of the grading operator, and
its concrete form can be computed by making use of the relations
presented above. Again, we  get a closed non-linear superalgebra,
the nonlinearity of which originates from the polynomial dependence
of the superalgebraic   structure functions on the  Hamiltonian
$\mathcal{H}^{PT}$, which plays a role of the multiplicative central
charge. Since we have defined $\mathcal{Q}^{PTab}_{l}$ as the
$Q^{PTab}_{l/\lambda}$ taken with the same virtual parameter
$\tau_\lambda=\infty$, the set of the second order integrals
$\mathcal{Q}^{PTab}_{l}$ together with the Hamiltonian
$\mathcal{H}^{PT}$ form the closed nonlinear sub-superalgebra.

If some of the displacement parameters $\tau_a$ are taken to be
infinite, corresponding Hamiltonians $H_a^{PT}$ transform into those
of the free particle. In this case we loose the property of the
self-isospectrality of the chain, but the supersymmetric structure
is still present and can also be computed by employing the relations
discussed above.

\section{Discussion and outlook}

We investigated the unusual nonlinear  supersymmetric structure  of
the self-isospectral crystalline chains formed by the arbitrary
number $N\geq 2$ of mutually displaced periodic one-gap Lam\'e
systems, and of the associated non-periodic self-isospectral soliton
chains described by reflectionless P\"oschl-Teller Hamiltonians. It
is generated by the $N(N-1)$ integrals of motion which are the first
order differential matrix operators, by the same number of the
second order matrix integrals, and by the $N$ third order Lax
integrals. The supersymmetry admits distinct choices for the grading
operator, that classifies  these integrals as bosonic and fermionic
operators in different ways.
 For instance, for the example of the  $N=3$ chain from  Section
6, three choices of the grading operator indicated in  Eq.
(\ref{3G}) classify the complete set of $15$  nontrivial local
integrals of motion as, respectively, $7+8=(2+2+3)+(4+4+0)$,
$6+9=(0+6+0)+(6+0+3)$ and $6+9=(4+2+0)+(2+4+3)$ bosonic$+$fermionic
generators, where the first, second and third numbers in each
parentheses correspond to the numbers of the first, second and third
order differential matrix operators.
 In dependence on the chosen  grading
operator,   one of the third order integrals, $L$, is the bosonic
central charge, or  the fermionic supercharge to be a square root of
the matrix spectral polynomial of the $N$-chain~\footnote{
 As we noted at the very beginning, 
 this happens even in the case $N=1$ for the unextended
Lam\'e system (\ref{HLame}), which is described by a hidden,
bosonized supersymmetry  with the reflection $\mathcal{R}$
identified as the grading operator \cite{MP1,CP1}.}.
 In the latter, unlike the former,  case the spectral polynomial
appears explicitly just in the superalgebraic relations. This
reveals the identifying characteristic of the Lax integrals: they
recognize the band edge states in the spectrum of each Lam\'e
subsystem by annihilating them.
  Another peculiarity is that the set of all
the second order integrals of motion taken with the same virtual
parameter generates together with the Hamiltonian a nonlinear
sub-superalgebra.

The lowest $N$-fold degenerate energy level was chosen to be zero,
and the spectra of the self-isospectral chains described by the
second order matrix Hamiltonians do not depend on the values of the
displacement parameters. On the other hand, according to Eqs.
(\ref{A1}) and (\ref{vareps}), the spectra of the first order matrix
integrals $S^{ab}_l$  depend on the mutual shifts $\tau_{ab}$, and
blow up when $\tau_{ab}$ tends to zero (modulo the period in the
case of the crystalline chain). This indicates on another
possibility to interpret the systems by identifying a suitable
combination of the first order integrals as a Hamiltonian. For
instance, for $N=2$, one can treat $S^{12}_1$ as the integral
$\mathcal{H}_{(1)}$, or, for $N=2n$ we can choose
$\mathcal{H}_{(1)}=S^{12}_1+S^{34}_1+\ldots S^{2n-1\, 2n}_1$. The
lower index indicates that the Hamiltonian is of the first order
(Dirac) nature, which can be considered as a kind of Bogoliubov-de
Gennes Hamiltonian in the Andreev approximation. Such
$\mathcal{H}_{(1)}$ in the  $N=2$ case was considered, for instance,
in the physics of conducting  polymers \cite{SaxBish,ThiesJPA}, or
as a Hamiltonian that describes the kink-antikink crystal
\cite{Thies,Basar} (or, kink-antikink baryons in the non-periodic
limit case \cite{FeiZee}) in the Gross-Neveu model. Therefore, the
$N>2$ chains in such a reinterpretation with the first order
Hamiltonian would provide some generalization of the known $N=2$
models, in which spectral gaps are governed by the displacement
parameters of the corresponding second order chains. The interesting
peculiarity of such \emph{first order} systems is that they possess
the \emph{ own nonlinear supersymmetry}. Indeed, the operator
$\hat{\Gamma}_1=\mathcal{RT}\Gamma$, where $\Gamma$ is given by Eq.
(\ref{Gamma}), commutes with $\mathcal{H}_{(1)}$,   and  can be
identified as a grading operator for such a first order system. The
Lax operator $L$ anti-commutes with $\hat{\Gamma}_1$, and the latter
classifies $\mathcal{H}_{(1)}$ and $L$ as, respectively, bosonic and
fermionic generators. Since $L$ commutes with $\mathcal{H}_{(1)}$,
it can be considered as a fermionic supercharge, whose square, in
accordance with  Eqs. (\ref{A1}) and (\ref{BB5}), gives some
polynomial of order six in $\mathcal{H}_{(1)}$ {}\footnote{
 Such a nonlinear supersymmetry in the first order systems
$\mathcal{H}_{(1)}$  was discussed in \cite{PAN,PNPRD} for the
simplest case of $N=2$ chains; it appears particularly in the
twisting of carbon nanotubes, see \cite{JPtwist}.
 }. For $n>1$, the system
$\mathcal{H}_{(1)}$ has also other nontrivial integrals of
motion, see Eq. (\ref{B1}) with $l=1$. 
 Such a nonlinear supersymmetry in the first order system
$\mathcal{H}_{(1)}$ could not be revealed, however, with the choice
of the grading operator in the reflection-independent form
(\ref{Gamma}) which identifies the Lax integral $L$ as the bosonic
operator and $\mathcal{H}_{(1)}$ as the fermionic one.

Another interesting possibility for generalization of the results is
to identify some linear combination of the second order matrix
operators as a Hamiltonian, for instance, by taking
$\mathcal{H}_{(2)}=Q^{12}_{1/\lambda} +Q^{34}_{1/\lambda}+\ldots
Q^{2n-1\, 2n}_{1/\lambda}$ in the case of $N=2n$. The spectrum of
$\mathcal{H}_{(2)}$ like that of $\mathcal{H}_{(1)}$ depends on the
shift parameters, see (\ref{A3}). For $N=2$ or $N=4$, such a second
order matrix Hamiltonian has a nature to be similar to that of the
Hamiltonian in the physics of bilayer graphene \cite{KatsKlein}.
With respect to the grading operator $\hat{\Gamma}=\mathcal{RT}$,
the Hamiltonian $\mathcal{H}_{(2)}$ and the third order operator $L$
are, respectively, the bosonic and fermionic operators. The operator
$L$ as well as the operators $L^{12}_1$, $\ldots$, $L^{2n-1\,
2n}_1$, see Eq. (\ref{B3}) with $l=1$, are the supercharges of the
nonlinear supersymmetry of the system described by the unusual
second order matrix Hamiltonian $\mathcal{H}_{(2)}$.
 Notice a special role played by the reflection-dependent grading
operator $\hat{\Gamma}=\mathcal{RT}$ for revealing the
supersymmetric structure in the indicated unusual second order
system $\mathcal{H}_{(2)}$.

The peculiarity of the non-periodic case in comparison with the
periodic one is that the chain subsystems  can be related there,
particularly,  by the second order intertwiners (\ref{YAA}). As the
intermediate (virtual) system in this case, there appears a free
particle system. The latter can be treated  as the P\"oschl-Teller
system (\ref{HlimPT}),
$H_\lambda^{PT}=H_\lambda^{PT}(x+\tau_\lambda)$, displaced to
infinity, $\tau_\lambda\rightarrow \infty$.  It is due to such a
relation the Lax operator (\ref{PlimPT}) has a nature of a dressed
free particle momentum operator, and eigenstates of $H^{PT}$ can be
obtained by the Darboux transformation of the corresponding free
particle eigenstates. We have, unfortunately,  no such  simple
relation with a free particle in the periodic case.

We considered the case of self-isospectral Hermitian chains with
real displacements. The construction can be generalized for the case
of complex shift parameters. The corresponding supersymmetric
structure can be interesting then in the context of the physics of
PT-symmetric systems \cite{PT1,PT2}, 
 where, again, the discrete transformation operators, particularly
spatial reflection, prove to play a fundamental role \cite{CoPl}.

\vskip0.2cm

 \noindent \textbf{Acknowledgements.}
The work has been partially supported by
 FONDECYT Grant 1095027, Chile.
MP thanks the Benasque Center for Science for a stimulating
environment, and University of Valladolid for hospitality during the
initial stages of the work.


\end{document}